\documentclass[prx,twocolumn,amsmath,amssymb,notitlepage]{revtex4-1}
\usepackage{tikz}
\usepackage{circuitikz}
\usetikzlibrary{arrows,shapes.gates.logic.US,shapes.gates.logic.IEC,calc}
\usepackage{graphicx}
\usepackage{amssymb,dsfont}
\usepackage{amsfonts}
\usepackage{amsmath}
\usepackage{amsbsy}
\usepackage{natbib}
\usepackage{comment}
\usepackage{amssymb}
\usepackage{color}
\usepackage{float}
\DeclareMathAlphabet\mathbfcal{OMS}{cmsy}{b}{n}
\DeclareMathAlphabet{\mathpzc}{OT1}{pzc}{m}{it}
\usepackage[colorlinks=true, urlcolor=blue, citecolor=blue, linkcolor=blue]{hyperref}
\usepackage{bm}% bold math
\usepackage{bm}
\usepackage{bbold}

\renewcommand{\vec}[1]{\mbox{\boldmath$\mathrm{#1}$}}

\renewcommand{\vec}[1]{\mbox{\boldmath$\mathrm{#1}$}}
\newcommand{\be}{\begin{equation}}
\newcommand{\ee}{\end{equation}}
\newcommand{\ben}{\begin{eqnarray}}
\newcommand{\een}{\end{eqnarray}}
\def\vl{|}

\def\ur{{\uparrow}}
\def\dr{{\downarrow}}

\begin{document}

%\preprint{APS/123-QED}

\title{Superconducting Diode sensor}

\author{A. Sinner$^{1}$, X.-G. Wang$^{2}$, S. S. P. Parkin$^{3}$, A. Ernst$^{3,4}$, V. Dugaev$^{6}$, and L. Chotorlishvili$^6$}
\address{$^1$Institute of Physics, University of Opole, 45-052 Opole, Poland\\
$^2$ School of Physics and Electronics, Central South University, Changsha 410083, China \\
$^3$ Max Planck Institute of Microstructure Physics, Weinberg 2, D-06120 Halle, Germany\\
$^{4}$ Institute for Theoretical Physics, Johannes Kepler University, Altenberger Stra\ss e 69, 4040 Linz, Austria \\
$^6$ Department of Physics and Medical Engineering, Rzesz\'ow University of Technology, 35-959 Rzesz\'ow, Poland}
\date{\today}% It is always \today, today,
\begin{abstract} 

We study the superconducting Josephson junction diode operating via the magnetic field of skyrmions. Inspired by the near-field optical microscopy, we propose to partially screen the magnetic field and analyze part-by-part the magnetic texture of the skyrmion. The detected asymmetric supercurrent is influenced by the skyrmionic magnetic field and magnetic texture. 
This enables the Josephson junction diode to function as a hyperfine sensor and to read out the information about the morphology of the complex magnetic textures. The proposed setup opens a new avenue in magnetometry and represents an alternative to the technologies based on the nitrogen-vacancy centers.
\end{abstract}

\maketitle

\textit{\textbf{Introduction:}} 
Diode is a device which exhibits the non-reciprocal responses and transport properties. The asymmetric conductance of the diode applies not only to the electric current but also its sonic counterparts, the propagation of acoustic vibrations, and the rectification of the heat transport at the nano-level. Recently the Josephson diode effect has been discovered~\cite{ando2020observation,jeon2022zero,pal2022josephson,davydova2022universal,baumgartner2022supercurrent}. In a Josephson junction, a thin metal or dielectric barrier layer separates two superconductors. When the inversion symmetry is broken by an external magnetic field, the Cooper pairs acquire a finite center-of-mass momentum, which leads to the phase mismatch between the left- and right-propagating currents and correspondingly to the asymmetric transport across the junction. This is the underlying mechanism of the Josephson diode effect.

In conventional Josephson junctions near the critical temperature, the current-phase relations are typically sinusoidal $j(\varphi)=j_c\sin(\varphi)$, where $j_c$ is the critical current,  $\varphi=\varphi_1-\varphi_2$ is the difference between phases of the superconducting order parameters $\Delta_{1,2}=\Delta e^{i\varphi^{}_{1,2}}$. 
%The symmetry of the system plays an important role. 
In the systems with preserved time-reversal symmetry the current-phase relation is always antisymmetric $j(\varphi)=-j(-\varphi)$. By contrast, the broken time-reversal symmetry leads to the $\varphi_0$-type of the current-phase relations $j(\varphi)=j_c\sin(\varphi+\varphi_0)$, cf. \cite{PhysRevLett.101.107005, PhysRevLett.128.037001} for more details.
The deeper analysis of the symmetry properties of the Onsager's coefficients suggests, that the two-terminal resistance must be even with respect to the magnetic field. This argument also applies to the case of the linear magnetoresistance, when resistance $\varrho$ depends linearly on the magnetic field $\textbf{B}$. Therefore to preserve the symmetry properties of the Onsager's coefficients, one needs to introduce an extra chiral term $\chi^{L/R}$ in the expression of the resistance $\varrho^{L,R}(\textbf{k},\textbf{B})=\varrho\left(1+\chi^{L/R} \textbf{I}\cdot \textbf{B}\right)$~\cite{PhysRevLett.87.236602}
%. Here, $\varrho^{L,R}(\textbf{k},\textbf{B})$ are the resistances 
for the left and right currents respectively. Here, $\textbf{k}$ is the wave vector, $I$ the current and the different sign $\chi^L=-\chi^R$ is required by the parity reversal symmetry. 
The occurrence of the chiral resistance $\varrho^{L,R}(\textbf{k},\textbf{B})$ is the most essential consequence of the magnetochiral anisotropy. Recently, the experimental observation of the superconducting diode effect has been reported~\cite{ando2020observation}, which triggered a substantial theoretical interest in this phenomenon~\cite{jeon2022zero,pal2022josephson,davydova2022universal,baumgartner2022supercurrent}.
Although it has been known that the Josephson junction can sense small magnetic fields, only the effects of uniform constant external magnetic fields have been investigated that far. 

\begin{figure}[t]\label{Schematics}
\centering
\includegraphics[scale=.3]{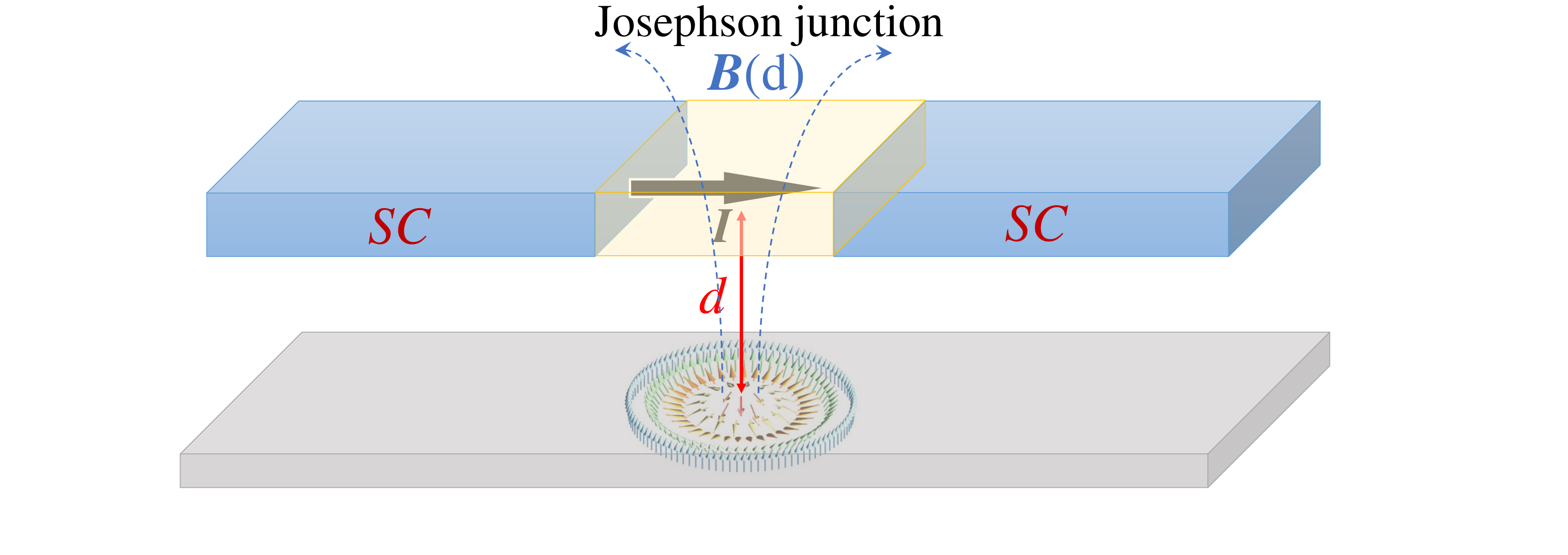}
\caption{The schematic representation of the Josephson junction sensoring device. The magnetic field of the skyrmion breaks the inversion symmetry of the Josephson junction and gives rise to the superconducting diode effect. To increase the sensibility of the sensor, one can partially cover the skyrmion surface and analyse the effect of the magnetic field from the uncovered part. It is  always possible to keep the open part of the skyrmion surface smaller than the characteristic size of the Josephson junction.
%The system's schematics shows that when the skyrmion's magnetic field crosses the Josephson junction, it disrupts the inversion symmetry, which results in the superconductor diode effect. In order to increase the sensibility of the superconducting Josephson junction sensor, one can cover part of the region of the skyrmion surface shield part of the magnetic lines and analyze the remaining uncovered part. Thus one can ensure that the uncovered region of the skyrmion texture is always smaller compared to the characteristic size of the Josephson junction.
%\vspace{2mm}
}
\label{fig:schematics}
\end{figure}

The superconducting diode implies the effect of the nonreciprocal charge transport, which can be achieved when both, the spatial inversion and the time-reversal symmetries are broken. The Rashba spin-orbit term breaks the uniaxial spatial inversion symmetry along the $\textbf{z}$-axis. 
The magnetochiral anisotropy is achieved in two realizations of external fields: 
(a) By applying the external magnetic field along the $\textbf{y}$-axis and the electric field along the $\textbf{x}$-axis. 
Then the non-reciprocal effect is quantified by the term 
$\varrho=\varrho_0\left[1+\gamma(\textbf{B}\times\textbf{z})\cdot\textbf{I}\right]$~\cite{ando2020observation};  
(b) Alternatively one can direct electric field along the $\textbf{z}$-axis. Then both, the magnetic field and the current are in-plane,
leading to the chiral resistance term $\varrho=\varrho_0\left[1+\gamma (\textbf{B}\times\textbf{I})\cdot \textbf{z}\right]$~\cite{baumgartner2022supercurrent}.
In previous studies, the magnetic field in the expression of chiral resistance has been associated with some homogeneous external field \cite{jeon2022zero,pal2022josephson,davydova2022universal,baumgartner2022supercurrent}. 
In contrast to that, in the present project we will consider the magnetic field generated by the complex magnetic textures, such as for example the skyrmionic magnetic textures. In particular, the proximity effect of the skyrmion lattice or of individual skyrmions and stray magnetic fields can violate the $\mathcal{PT}$-symmetry. The skyrmion exerts the stray magnetic field (odd in $\mathcal{T}$ and even in $\mathcal{P}$) on the Josephson junction and violates $\mathcal{PT}$-symmetry. The superconducting diode should in principle feel the stray magnetic field and read out the information about the magnetic texture Fig.(\ref{fig:schematics}). In the present project, we propose the superfine superconducting diode sensor. Such a sensor can open new avenues in experimental skyrmionics, e.g. for studying the morphology of complex magnetic textures and systems. 
In particular, the superconducting diode sensors would be able to identify different magnetic phases, such as Néel and Bloch skyrmions, or ferromagnetic phases, and monitor transitions between them in a time domain. 

In order to increase the sensibility of the superconducting Josephson junction sensor, we need to analyze the magnetic field (i.e., skyrmion magnetic texture) part by part separately for different flakes of the skyrmion texture. Thus we are looking for the spatially resolved effect. Near-field optical microscopy is a technology that allows covering the surface from unwanted electromagnetic interactions except for the small selected regions of 20-30nm in characteristic size. The magnetic field lines are closed loops. However, magnetic lines also can be redirected by offering them a preferred path (e.g., using high permeability materials to shield the magnetostatic field) \cite{grabchikov2016effectiveness}. Recent studies show that graphene can serve as a perfect shield due to its extraordinary properties~\cite{chen2021recent}.

The expression for the superconducting current can be derived phenomenologically~\cite{jeon2022zero,davydova2022universal}. 
The free energy of the system has the form
\begin{eqnarray}
\label{eq:free.energy}
F=-2\vert\gamma_1\vert\Delta^2\cos\varphi-\vert\gamma_2\vert\Delta^4\cos(2\varphi+\delta),
\end{eqnarray} 
where $\gamma_1$ and $\gamma_2$ denote the first- and second-order Cooper pair tunnelling processes. 
We derive the explicit expressions for $\gamma_{1,2}(\textbf{B})$ from the microscopic theory. 
The key issue is the explicit dependence of coefficients $\gamma_{1,2}$ on the magnetic field. 
The phases in Eq.~(\ref{eq:free.energy}) are defined as follows: 
$\varphi=\varphi_2-\varphi_1+\text{arg}(\gamma_1)$, $\delta=\text{arg}(\gamma_2)-2\text{arg}(\gamma_1)$. 
The superconducting current $I\left(\varphi\right)=\frac{2\pi}{\Phi_0}\frac{\partial F}{\partial \varphi}$ becomes
\begin{eqnarray}
\label{superconducting current}
I\left(\varphi\right)=\frac{2\pi}{\Phi_0}\left\lbrace \Delta^2\vert\gamma_1\vert\sin\varphi+\Delta^4\vert\gamma_2\vert\sin(2\varphi+\delta)\right\rbrace, 
\end{eqnarray}
where $\Phi_0={h}/{e}$ denotes the superconducting flux quantum, $e$ the charge of the electron. 
From Eq.(\ref{superconducting current}) it is easy to see that the asymmetry between the left and right currents arises due to the term proportional to $\Delta^4$.
%\begin{equation}
%\label{eq:Current_asymmetry}
%\delta I(\varphi)=I(\varphi)-\vert I(-\varphi)\vert. 
%\end{equation}
The skyrmions emerge in the thin magnetic films with broken inversion symmetry. 
The lack of inversion symmetry is associated with the Dzyaloshinskii–Moriya interaction (DMI) \cite{white2014electric,derras2018quantum, haldar2018first,psaroudaki2017quantum,van2013magnetic,rohart2016path,tsesses2018optical,wang2018electric,wang2019thermally,PhysRevB.106.104424,PhysRevB.107.094404,PhysRevB.107.L100419}. 
Below we describe the magnetic texture of skyrmions by the local magnetization $\textbf{M}(\textbf{r})$.
The dynamics of the local magnetization is governed by the phenomenological Landau-Lifshitz-Gilbert (LLG) equation
\begin{equation}
\displaystyle \frac{\partial \vec{M}}{\partial t} = - \gamma \vec{M} \times \vec{H}_{\mathrm{\rm eff}} + \frac{\alpha}{M_{s}} \vec{M} \times \frac{\partial \vec{M}}{\partial t}.
\label{LLG}
\end{equation}
Here, $ \vec{M}(\textbf{r}) = M_s \vec{m} $ describes the magnetic texture of the thin magnetic film, $ M_s$ being the saturation magnetization, $\vec{m}$ is the unit vector along the magnetization direction $\vec{M}$, and $ \alpha $ is the phenomenological Gilbert damping constant. 
The total effective magnetic field $ \vec{H}_{eff}$ reads
$\displaystyle\vec{H}_{eff}= \frac{2 A_{ex}}{\mu_0 M_{s}} \nabla^2 \vec{m} + H_z
\vec{z}  -\frac{1}{\mu_0 M_s} \frac{\delta E_D}{\delta
 \vec{m}} $,
where the first term describes the internal exchange field with the
exchange stiffness $ A_{ex} $, the second term corresponds to the
external magnetic field $ H_z $ ($\vec{z}$ is a unit vector along the
axis $z$ normal to the film),  while the last term is the DM
field, with the DM interaction energy density
$ E_D = D_m [(m_z \frac{d m_x}{dx} - m_x \frac{dm_z}{dx}) + (m_z
\frac{d m_y}{dy} - m_y \frac{dm_z}{dy})] $
and $ D_m $ being the strength of the DM interaction. 

For numerical calculations we use the set of parameters which correspond to Co/heavy-metal multi-layers:
$ A_{ex} = 10 \rm{pJ}/\rm{m} $, $ D_m = 0.2 \rm{mJ}/\rm{m}^{2} $, and
$ M_s = 1.2\times 10^6 \rm{A}/\rm{m}$.  The bias magnetic field
$ H_z = 100 $ mT is used for stabilization of the skyrmion
structure. The skyrmion width is 45 nm. The size of the ferromagnetic layer is
$ 3000 \times 240 \times 3 \rm{nm}^3$. The ferromagnetic layer is discretized with the cells of size
$ 3 \times 3 \times 3 \rm{nm}^3 $.
Of particular importance is structure of the stray magnetic field $\textbf{B}(\textbf{m}(\textbf{r}))$ generated by the magnetic skyrmion texture $\vec{m}(\textbf{r})$. The explicit expression for the stray magnetic field can be derived analytically \cite{dovzhenko2018magnetostatic}, but the derivation is rather involved and is placed into the Supplementary Information~\cite{SI}. 

For the microscopic estimation of the parameters of the phenomenological free-energy Eq.(\ref{eq:free.energy}) 
we exploit the Bogoliubov-deGennes Hamiltonian of the junction 
\begin{equation}
\label{eq:BdGH}
{\rm H}^{}_{\rm BdG} = \left( 
 \begin{array}{cccc}
 h^{}_0 - \mu & {\bm\Delta}^\ast_{0,1} &  {\bm t}^\ast &  0 \\
 {\bm\Delta}^{}_{0,1} & - h^{}_0 + \mu &  0 & {\bm t}^{\ast} \\
 {\bm t} &  0 & h^{}_0 - \mu & {\bm\Delta}^\ast_{0,2} \\
  0 & {\bm t}  &  {\bm \Delta}^{}_{0,2} &  - h^{}_0 + \mu
 \end{array}
 \right),
\end{equation}
where $\mu$ denotes the chemical potential, which is applied to both sides of the junction and $h^{}_0 = -\frac{\hbar^2}{2m}\nabla^2$.
$ {\bm\Delta}^{}_{0,i}= \Delta^{}_{0}e^{i\chi^{}_i}$, $i=1,2$ denotes the mean-field order parameter in each superconductor of the junction. 
Furthermore, ${\bm t}=te^{i\xi}$ is the complex tunneling amplitude across the junction, related to the overlap of the electronic 
quantum wave functions on each superconducting side. Here we do not consider the tunneling matrix elements in the spin-flip channel,
since flipping spins costs additional energy. To set up the free-energy energy functional, we have to consider the fluctuations in the order parameter above its mean-field value. In the simplest approximation, the fluctuation term would read~\cite{PhysRevResearch.2.033085}
\begin{equation}
\label{eq:FlH}
\delta{\rm H} = 
\left( 
 \begin{array}{cccc}
 0 &  {\bm\Delta}^\ast_{1} & 0 &  0 \\
  {\bm\Delta}^{}_{1} &  0 &  0 & 0 \\
  0 &  0 & 0 &  {\bm\Delta}^\ast_{2} \\
  0 & 0 &   {\bm\Delta}^{}_{2} & 0
 \end{array}
\right),
\end{equation}
where ${\bm\Delta}^{}_{1,2} = \Delta e^{i\varphi^{}_{1,2}}$ are the time and position-dependent fluctuations of the order parameters corresponding to each superconductor. 
The free energy $\cal F$ is related to the grand-canonical partition function $\cal Z$ via 
\begin{equation}
{\cal Z} = e^{-{\cal F}},
\end{equation}
where the properly normalized zero-temperature grand canonical partition function reads 
\begin{equation}
\label{eq:GCPF}
{\cal Z}[\delta{\rm H}] = \lim_{\beta\to\infty}\frac{\displaystyle{\rm Tr}~e^{-\beta({\rm H}^{}_{\rm BdG}+\delta{\rm H})}}
{\displaystyle{\rm Tr}~e^{-\beta{\rm H}^{}_{\rm BdG}}},
\end{equation}
and $\beta=(k^{}_B T)^{-1}$. 
Within the functional integrals formalism, the grand-canonical partition function Eq.~(\ref{eq:GCPF}) becomes
\begin{equation}
\label{eq:Func-Int}
{\cal Z}[\delta{\rm H}] = \frac{1}{{\cal Z}^{}_0}\int{\cal D}\bar\Psi{\cal D}\Psi ~ e^{-{\cal S}[\bar\Psi,\Psi]} ,
% =  \exp{\rm tr}\log[\mathds{1} + G^{}_0\delta{\rm H}],
\end{equation}
where ${\cal Z}^{}_0={\cal Z}[0]$. The corresponding action comprises two terms ${\cal S} = {\cal S}^{}_{\rm BdG} + \delta{\cal S} $, where the first part corresponds to the mean-field Hamiltonian and the second to the term containing the fluctuations of the order parameters. Concretely we have ${\cal S}^{}_{\rm BdG} = \bar\Psi\cdot[\hbar\partial^{}_\tau{\mathds 1} + {\rm H}^{}_{\rm BdG}]\Psi$, $\delta{\cal S} = \bar\Psi\cdot\delta{\rm H}\Psi$ with ${\rm H}^{}_{\rm BdG}$ and $\delta{\rm H}$ defined in Eq.~(\ref{eq:BdGH}) and Eq.~(\ref{eq:FlH}). 
The Grassmann fields $\psi^\dag,\psi$, and $\varphi^\dag,\varphi$ corresponding to both leads of the junction are combined to the Nambu bispinors $\bar\Psi = (\psi^\dag_\ur,\psi^{}_\dr,\varphi^\dag_\ur,\varphi^{}_\dr)$, where the up- and down-arrows denote two spin projections. The action is quadratic in Grassmann fields, and their integration can be performed exactly, which leads us to the general expression of the free energy as a function of fluctuations. The evaluation of this expression is conveniently performed by functional integrals, which leads us to  
\begin{equation}
{\cal F}[\delta{\rm H}] = -{\rm tr}\log[\mathds{1} + G^{}_0\delta{\rm H}],
\end{equation}
with the imaginary time Green's function 
\begin{equation}
G^{}_{0}(r\tau;r^\prime\tau^\prime) = 
\langle r,\tau\vl[\partial^{}_\tau{\mathds 1} + {\rm H}^{}_{\rm BdG}]^{-1}\vl r^\prime \tau^\prime\rangle.
\end{equation}
In the expansion in powers of $\delta{\rm H}$, we only keep terms which are present in the effective action Eq.(\ref{eq:free.energy}). 
To the second order we get $\frac{1}{2}{\rm tr}(G^{}_0\delta{\rm H})^2 \approx  2\gamma^{}_1 \Delta^{}_1\Delta^\ast_2 + 2\gamma^\ast_1 \Delta^\ast_1\Delta^{}_2 + \cdots$, with the coupling parameter
\begin{equation}
\label{eq:2ndOrCoupling}
\gamma^{}_1 \approx e^{-i(\chi^{}_1 - \chi^{}_2)} \frac{1}{2}\frac{\rho(E^{}_F) t^2}{3!\Delta^2_0} \cos\gamma,
\end{equation}
where $\gamma = 2\xi - \chi^{}_1 + \chi^{}_2$.  
The phases of the type $\gamma$ often appear in multipartite superconducting systems, e.g.~\cite{PhysRevB.104.245124}.
The density of states (DOS) for two spin projections of the 3d free electron gas at the Fermi surface reads 
\begin{equation}\label{in SI units}
\rho^{}_0(E^{}_F) = \frac{1}{2\pi^2}\left(\frac{2m}{\hbar}\right)^{\frac{3}{2}}\frac{\sqrt{E^{}_F}}{\Omega^{}_{\rm BZ}},
\end{equation}
where $\Omega^{}_{\rm BZ}$ denotes the volume of the Brillouin zone. Hence, $\rho^{}_0(E^{}_F)$ and correspondingly $\gamma^{}_1$ have the units of inverse energy. From the fourth order expansion term
$\frac{1}{4}{\rm tr}(G^{}_0\delta{\rm H})^4 \approx  \gamma^{}_2 (\Delta^{}_1\Delta^\ast_2)^2 + \gamma^\ast_2 (\Delta^\ast_1\Delta^{}_2)^2 + \cdots$ we extract the explicit form of the coupling parameter $\gamma_2$ as follows:
\begin{equation}
\label{eq:4thOrCoupling}
\gamma^{}_2 \approx e^{-i(2\chi^{}_1 - 2\chi^{}_2+\phi[\gamma])}\frac{\rho(E^{}_F) t^2}{5!\Delta^4_0} g[\gamma],
\end{equation}
which has the dimension of ${\rm energy}^{-3}$. Here we have introduced $g[\gamma] = \sqrt{49\cos^2\gamma + 25\sin^2\gamma}$ and $\phi[\gamma] = {\rm atan}\left[\frac{5}{7}\tan\gamma\right]$. Note that in general, in Eqs.~(\ref{eq:2ndOrCoupling}) and (\ref{eq:4thOrCoupling}) the fluctuations depend on both time and spatial coordinates and the summation over these variables has to be understood. 
With Eq.~(\ref{eq:2ndOrCoupling}) and (\ref{eq:4thOrCoupling}) follows for the time-reversal symmetry breaking phase shift $\delta=\text{arg}(\gamma_2)-2\text{arg}(\gamma_1) = \phi[\gamma]$. The diode current across the junction is proportional to $\sin[\delta]$, which gives $\delta I\left(\pm\frac{\pi}{2}\right) \sim \sin[\delta] \sim \sin[\gamma]$. Hence, the superconducting diode effect is rather general and in principle should be observable in any junction even without an external magnetic field. However, it is to expect that without an external field, the  
angle $\gamma$ is very small and the diode current would be suppressed by the thermal or mechanical noise. 
In the presence of the magnetic field, the phase difference is fixed to 
$\delta = 2\frac{e}{\hbar}\oint_{\cal C} d\vec r\cdot\vec A$ with the vector potential $\vec A$ related to the magnetic field via $\vec B$=\textbf{curl}$\vec A$ and factor 2 counting two superconductors on both sides of the junction~\cite{SI}. Finally we use the Stokes theorem $\oint_{\cal C} d\vec r\cdot\vec A = \int_{S} d\vec S\cdot{\bf curl}\vec A = \int_{S} d\vec S\cdot\vec B$. For instance this yields for the homogeneous y-directed magnetic field~\cite{SI}
\begin{equation}\label{eq:for the homogeneous field}
\delta = 2B^{}_y\frac{e}{\hbar}\int^h_{0}dz\int^{\lambda^{}_L}_0dx = \frac{B^{}_y}{B^{}_d},
\end{equation}
where $h$ is the height of the superconducting lead and $\lambda^{}_L$ the London penetration length. Introducing the flux $\Phi$ we can further write $\delta = h\lambda^{}_L \frac{e}{\hbar}B^{}_y = 2\pi\frac{\Phi}{\Phi^{}_0}$. 
Eq.~(\ref{eq:for the homogeneous field}) recovers the earlier result of the work~\cite{davydova2022universal}.

\begin{figure}[t]
\centering
\includegraphics[width=8cm]{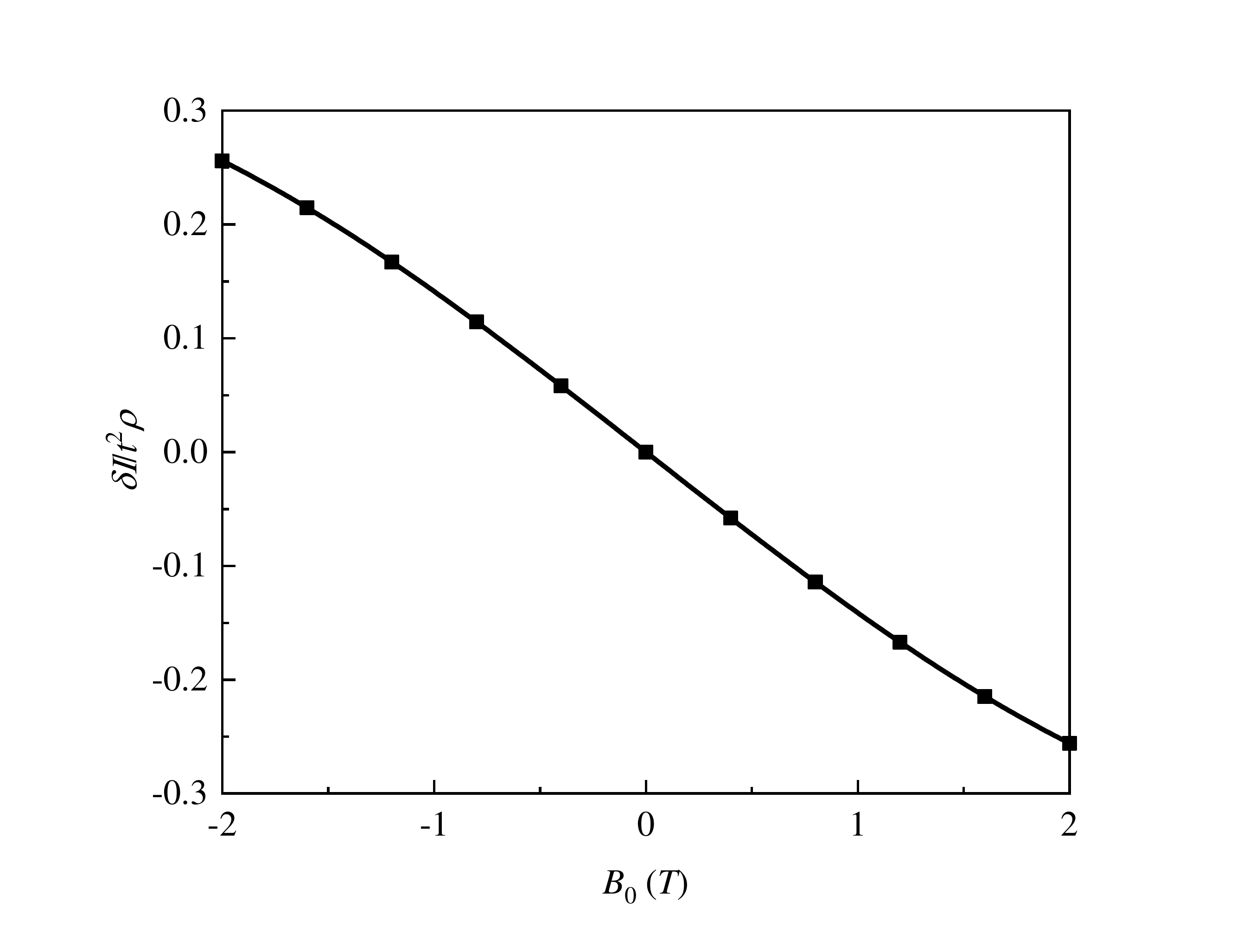}
\caption{The dependence of the supercurrent asymmetry $\frac{\delta I}{t^2\rho(E_F, B)}$ on homogeneous external magnetic field. Varying the external magnetic field from $B_0<0$ to $B_0>0$ we observe the change of the sign of current asymmetry}
\label{fig:magnet}
\end{figure}

\begin{figure}[t]
\includegraphics[width=8.5cm]{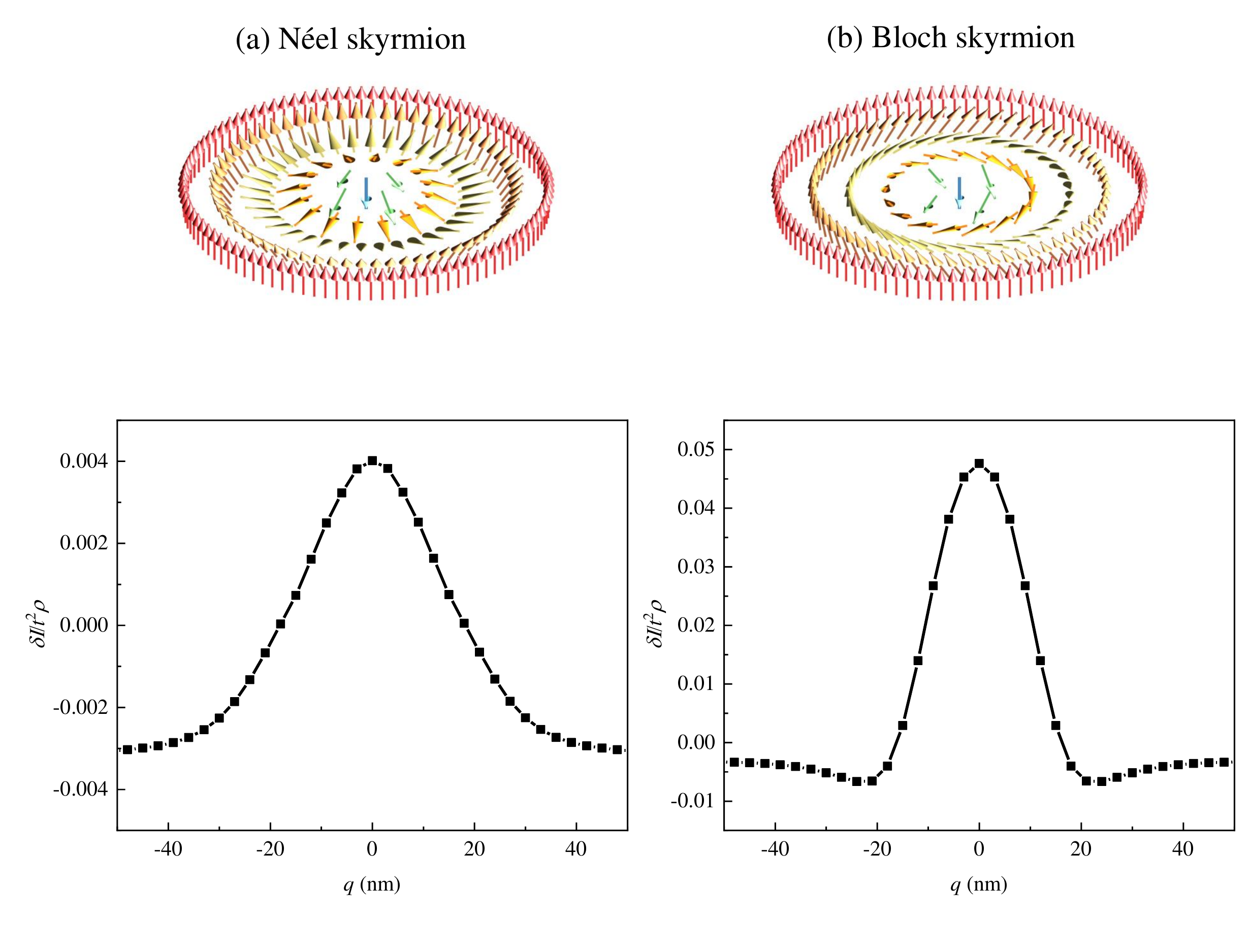}
%\hspace{10mm}
\includegraphics[width=7cm]{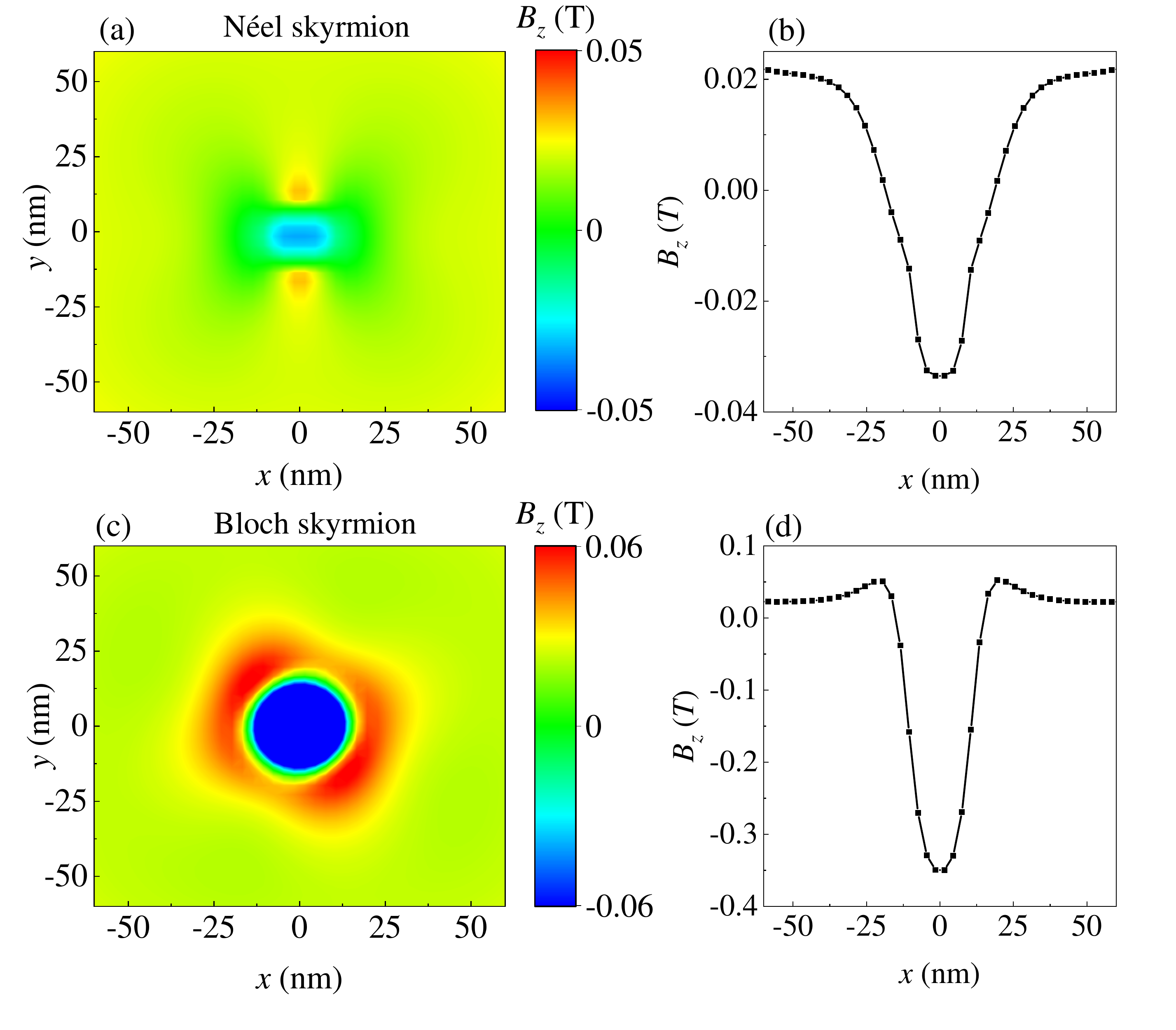}
\caption{
{\bf Top:} Variation of the supercurrent asymmetry in an inhomogeneous skyrmion field for (a) N\'eel  and (b) Bloch skyrmions. $q$ is the distance from the unscreened region to the center of the skyrmion at $(x=0, y=0)$. 
{\bf Bottom:} Skyrmion magnetic field component $B_z(x,y)$ as function of the distance from the skyrmion center  $(x=0, y=0)$. (b), (d) Crosssection profiles of the magnetic field components  $B_z(x,y=0)$ for (b) N\'eel  and  (d) Bloch skyrmions.
The distance between the skyrmion and the diode is $d=1.5$nm for all figures.
}
\label{fig:Syrmion}
\end{figure}

%\begin{figure}[t]
%\centering
%\includegraphics[width=7cm]{Fig.3.pdf}
%\caption{The change of the current asymmetry sign in the inhomogeneous skyrmion field for the (a) Neel and (b) Bloch skyrmions. $q$ is the distance from the unscreened region to the center of the skyrmion $(x=0, y=0)$. The distance between the skyrmion and the diode is $d=1.5$nm.}
%$$\label{fig:Neel and Bloch}
%$\end{figure}
%$$\begin{figure}[t]
%\centering
%\includegraphics[width=7cm]{Fig.4.pdf}
%\caption{Skyrmion magnetic field component $B_z(x,y)$ as a function of the distance from the skyrmion center  $(x=0, y=0)$. The %distance between the skyrmion and the diode is $d=1.5$nm. (b), (d) Section of the profiles of the magnetic field components  %$B_z(x,y=0)$ for Neel (b) and  Bloch (d) skyrmions.}
%\label{fig:Non-uniform and current}
%\end{figure}

The magnetic stray field $\textbf{B}(\textbf{r})$ induced by the skyrmion does not only influence the phases of the condensate but also the coupling parameters of the free energy, entering it via the density of the states. The complexity of our problem is due to the fact that the magnetic stray field is not uniform, and the corresponding Hamiltonian  $\hat h(\vec B) = -\frac{\hbar^2}{2m}\nabla^{2}_r \sigma^{}_0 + \mu^{}_B \vec\sigma\cdot\vec B(\textbf{r})$, with $\mu^{}_B=e\hbar/(2m)\approx 5.63\cdot10^{-5}$eV/T being the Bohr's magneton and the Lande factor assumed to be 2, cannot be generally diagonalized exactly. Thus the problem in our case is much more demanding than in earlier studies. 
The cumbersome evaluation and the final results for the DOS are presented in the Supplementary Information~\cite{SI}.

The effect of the magnetic field on the DOS arises from both, the topography (i.e. N\'eel or Bloch skyrmion) and the geometry (size) of the magnetic texture. Our aim is to explore these parameters via the rectification of the superconducting diode current. The details of calculations are presented in the Supplementary Information~\cite{SI}. 
The resulting supercurrent asymmetry reads
\begin{equation}
\label{eq:Asymmetrytext2}
\delta I = -t^2\rho(E_F,\vec B)\frac{|\Delta|^4}{3\Delta_0^4} \sin[\delta].
\end{equation}

We calculate the phase difference $\delta=\chi^{}_2 - \chi^{}_1$ with the Stokes' theorem $\oint_{\cal C} d\vec r\cdot\vec A = \int_{S} d\vec S\cdot\vec B $. 
Taking account of the particular device geometry we acquire:
\begin{equation}
\delta = \frac{4\pi}{\phi^{}_0} \intop^{\lambda^{}_L}_{0}dx~\intop^{\frac{W}{2}}_{-\frac{W}{2}} dy ~ B^{}_z,
\end{equation}
where $\lambda_L$ is the London penetration length. 
The explicit form of the skyrmion field $B_z$ is given in~\cite{SI}.

\textit{\textbf{Discussions}}:  
The skyrmion magnetic field enters the supercurrent asymmetry Eq.~(\ref{eq:Asymmetrytext2}) via the DOS and the phase factor. First we study the dependence of the supercurrent asymmetry $\frac{\delta I}{t^2\rho(E_F,\vec B)}$ on the external homogeneous magnetic field. Varying the values of the external magnetic field from negative $B_0<0$ to positive $B_0>0$ value, we observe the change of the sign of current asymmetry shown in the left panel of Fig.~\ref{fig:magnet}. A similar behavior is expected to be seen in the case of the inhomogeneous field of a skyrmion. To increase the sensitivity of the diode sensor, we screen most of the surface of the skyrmion and leave the small flake region uncovered to contribute to the effect. The center of the unscreened square region with the size $6\times6$ nm is spatially separated from the skyrmion center at $(x=0, y=0)$ by a characteristic distance $q$. It makes the superconducting diode a hyperfine spatially resolved sensor. The amplitude and the spatially resolved profile of the asymmetry rectification effect depend on the type of skyrmion and magnetic texture. This is shown in the upper panel in Fig.~\ref{fig:Syrmion}. Changing of the sign of $\delta I$ we can vary the sign of the corresponding spatially resolved magnetic field and the magnetic texture, cf. the bottom panel of Fig.~\ref{fig:Syrmion}.

\textit{\textbf{Summary}}:  
In this paper we develop the phenomenological and theoretical framework of the hyperfine sensing device based on the
superconducting Josephson diode effect. The main purpose of the device is to investigate the structure of the magnetic films textures with special emphasis on the high precision skyrmion morphology and location detection. By screening off large areas of the films, e.g. by graphene microshields, the effective exit area of the magnetic field is reduced to the scales comparable to the typical size of the Josephson junction. The sensing process consists in tracking the supercurrent anysotropy, which is influenced by the magnetic field of the film. Our theoretical modelling demonstrates persuasively that the supercurrent anysotropy imitates the topography of the investigated magnetic tissue of the film.

\textbf{\textit{Acknowledgements}}: The work is supported by Shota Rustaveli National Science Foundation of Georgia (SRNSFG) (Grant No. FR-19-4049), the National Natural Science Foundation of China (Grants No. 12174452, No. 11704415 and No. 12074437), the Natural Science Foundation of Hunan Province of China (Grants No. 2022JJ20050 and No. 2021JJ30784). This work was also supported by the National Science Center in Poland as a research project No. DEC-2017/27/B/ST3/02881.

%\bibliography{2nems}

%

\end{document}

%%%%%%%%%%%%%%%% SUPPLEMENTARY INFORMATION %%%%%%%%%%%%%%%%%%%%%%

\title{Supplementary Material \textit{for} \\Superconducting Diode sensor}

\author{A. Sinner$^{1}$, X.-G. Wang$^{2}$, S. S. P. Parkin$^{3}$, A. Ernst$^{3,4}$, V. Dugaev$^{6}$, and L. Chotorlishvili$^{6}$}
\address{$^1$Institute of Physics, University of Opole, 45-052 Opole, Poland\\
$^2$ School of Physics and Electronics, Central South University, Changsha 410083, China \\
$^3$ Max Planck Institute of Microstructure Physics, Weinberg 2, D-06120 Halle, Germany\\
$^{4}$ Institute for Theoretical Physics, Johannes Kepler University, Altenberger Stra\ss e 69, 4040 Linz, Austria \\
$^{6}$ Department of Physics and Medical Engineering, Rzesz\'ow University of Technology, 35-959 Rzesz\'ow, Poland}
%\date{\today}
\maketitle
 
%\section{Supplementary information}

\section{Derivation of the coupling parameters of the free energy functional}

Integration of the Nambu-fermions in the partition function Eq.(\textcolor{blue}{8}) in the main text leads us to the general expression for the free energy as function of fluctuations 
\begin{equation}
{\cal F}[\delta{\rm H}] = -\log\left[\frac{{\rm det}[G^{-1}_0+\delta{\rm H}]}{{\rm det}[G^{-1}_0]}\right] =  -{\rm tr}\log[\mathds{1} + G^{}_0\delta{\rm H}].
\end{equation}
Expansion to the fourth order in the fluctuations $G^{}_0\delta{\rm H}$ generates each term from the phenomenological free energy 
Eq.~(\textcolor{blue}{1}) in the main text, but also additional ones, which the minimal ansatz Eq.~(\textcolor{blue}{1}) in the main text does not suggest
\begin{equation}
{\cal F}[\delta{\rm H}] \approx - {\rm tr}\left[(G^{}_0\delta{\rm H}) - \frac{1}{2}(G^{}_0\delta{\rm H})^2 + \frac{1}{3}(G^{}_0\delta{\rm H})^3 - \frac{1}{4}(G^{}_0\delta{\rm H})^4 + \cdots  \right].
\end{equation}
To estimate the coupling parameters $\gamma^{}_1$ and $\gamma^{}_2$ it is sufficient to evaluate each loop to leading order in the hopping amplitude at zero temperature. Then the summation over Matsubara frequences becomes an integral stretching from $-\infty$ to $+\infty$.
First order loop cannot produce terms which mix BCS gaps from both leads. Therefore we do not consider it here. 
Second order loop to zeroth order in gradient expansion becomes
\begin{equation}
\frac{1}{2}{\rm tr}(G^{}_0\delta{\rm H})^2 \approx \frac{1}{2}{\rm Tr}\intop dX\int^{\infty}_{-\infty}\frac{d\omega}{2\pi} \int\frac{d^dk}{(2\pi)^d}~
G^{}_0(K)\delta{\rm H}^{}_XG^{}_0(K)\delta{\rm H}^{}_X.
\end{equation}
The shorthands introduced here are $X=(\tau,\vec x)$, $\tau$ being the imaginary time and $\vec x$ the $d$-dimensional vector in the position space,
and $K=(\omega^{}_n,\vec k)$, comprising of bosonic Matsubara frequency $\omega^{}_n$ and $d$-dimensional momentum vector $\vec k$. 
The expression under the integral depends on momentum $k$ only through the energy $h^{}_0=k^2/2m - \mu$, which enables us to perform the usual 
trick
\begin{equation}
\int\frac{d^dk}{(2\pi)^d}~F(h^{}_0) = \int\frac{d^dk}{(2\pi)^d}~\underbrace{\intop^{\infty}_{0}d\epsilon~\delta(\epsilon-h^{}_0)}_{=1} F(h^{}_0) = 
\intop^{\infty}_{0}d\epsilon~ F(\epsilon)~\underbrace{\intop\frac{d^dk}{(2\pi)^d}~\delta(\epsilon-h^{}_0)}_{\rm density~of~ states}
\approx \rho^{}_0(E^{}_F) \intop^{\infty}_{0}d\epsilon~ F(\epsilon),
\end{equation}
where $\rho^{}_0(E^{}_F)$ is the density of states of the non-interacting Fermi gas at the Fermi surface. 
For the terms which mix order parameters from both leads, the expansion to the leading order in $t$ gives 
\begin{equation}
\frac{1}{2}{\rm tr}(G^{}_0\delta{\rm H})^2 \approx 2\gamma^{}_1 \Delta^{}_1\Delta^\ast_2 + 2\gamma^\ast_1 \Delta^\ast_1\Delta^{}_2 + 
 2\tau^{}_1 \Delta^{}_1\Delta^{}_2 + 2\tau^\ast_1 \Delta^\ast_1\Delta^\ast_2 , 
\end{equation}
where 
\begin{equation}
\label{eq:2ndOrCouplings}
\gamma^{}_1 \approx e^{-i(\chi^{}_1 - \chi^{}_2)}\rho^{}_0(E^{}_F)\frac{1}{2} \frac{t^2}{3!\Delta^2_0} \cos[2\xi - \chi^{}_1 + \chi^{}_2] , \;\;\;
\tau^{}_1 \approx e^{-i(\chi^{}_1 + \chi^{}_2)}\rho^{}_0(E^{}_F)\frac{1}{2}\frac{t^2}{3!\Delta^2_0} \cos[2\xi - \chi^{}_1 + \chi^{}_2].
\end{equation}
We notice the phase shift between the phases of the "normal" ($\gamma^{}_1$, i.e. ${\rm arg}(\gamma^{}_1)=\chi^{}_1 - \chi^{}_2$) and "anomalous" ($\tau^{}_1$, i.e. ${\rm arg}(\tau^{}_1)=\chi^{}_1 + \chi^{}_2$) terms. In particular, in uniform magnetic fields there $\chi^{}_1 = -\chi^{}_2$,  
${\rm arg}(\gamma^{}_1)=0$ and therefore this term does not contribute to the current asymmetry.

The cubic term $\displaystyle -{\rm tr}(G^{}_0\delta{\rm H})^3/{3}$ has no corresponding term in the minimal phenomenological free-energy ansatz Eq.~(\textcolor{blue}{6}) in the main text and we skip the discussion of this term here. Quatric terms read to the zeroth order in gradient expansion 
\begin{equation}
\frac{1}{4}{\rm tr}(G^{}_0\delta{\rm H})^4 \approx \frac{1}{4}{\rm Tr}
\int dX\int dY\int dZ \int dR 
\intop^{\infty}_{-\infty}\frac{d\omega}{2\pi} \int\frac{d^dk}{(2\pi)^d}~
G^{}_0(K)\delta{\rm H}^{}_X G^{}_0(K)\delta{\rm H}^{}_Y G^{}_0(K)\delta{\rm H}^{}_Z G^{}_0(K)\delta{\rm H}^{}_R.
\end{equation}
We get
\begin{eqnarray}
\frac{1}{4}{\rm tr}(G^{}_0\delta{\rm H})^4 &\approx& \frac{\rho t^2}{5! \Delta^4_0}\left\{
(\Delta^{}_1\Delta^\ast_2)^2~e^{-2i(\chi^{}_1-\chi^{}_2)}\left(7\cos[2\xi-\chi^{}_1+\chi^{}_2] + 5i \sin[2\xi-\chi^{}_1+\chi^{}_2]\right) \right. \\
&+&(\Delta^\ast_1\Delta^{}_2)^2~e^{2i(\chi^{}_1-\chi^{}_2)}\left(7\cos[2\xi-\chi^{}_1+\chi^{}_2] - 5i \sin[2\xi-\chi^{}_1+\chi^{}_2]\right) + 
\cdots\left.\right\}.
\end{eqnarray}
Introducing the polar notation for the complex numbers we bring the coupling parameter to the form used in Eq.~(\textcolor{blue}{13}).

%\newpage

\section{Free energy functional in microscopic and phenomenological approaches}

We will use the following notation 
\begin{equation}
\gamma^{}_1 = e^{i{\rm arg}(\gamma^{}_1)}\vl\gamma^{}_1\vl,\;\; \gamma^{}_2 = e^{i{\rm arg}(\gamma^{}_2)}\vl\gamma^{}_2\vl,\;\; 
\Delta^{}_1 = e^{i\varphi^{}_1}\vl\Delta\vl,\;\; \Delta^{}_2 = e^{i\varphi^{}_2}\vl\Delta\vl.
\end{equation}
The phases ${\rm arg}(\gamma^{}_{1,2})$ are read off Eqs.~(\textcolor{blue}{11}) and (\textcolor{blue}{13}) in the main text:
\begin{equation}
\label{eq:PhasesOfGammas}
 {\rm arg}(\gamma^{}_{1}) = \chi^{}_1 - \chi^{}_2,
 \;\;
 {\rm arg}(\gamma^{}_{2}) = 2\chi^{}_1 - 2\chi^{}_2+\phi[\gamma],
\end{equation}
with $\phi[\gamma]$ being the phase shift, which occurs in the fourth order term of fluctuation expansion and is defined in the text under Eqs.~(\textcolor{blue}{13}) in the main text. Then the free energy functional in phenomenological truncation \cite{pal2022josephson,davydova2022universal} reads
\begin{eqnarray}
{\cal F} &=& -2\cos[ {\rm arg}(\gamma^{}_{1}) + \varphi^{}_1 - \varphi^{}_2] \vl\gamma^{}_1\vl\vl\Delta\vl^2 - 
2\cos[ {\rm arg}(\gamma^{}_{2}) + 2\varphi^{}_1 - 2\varphi^{}_2]\vl\gamma^{}_2\vl\vl\Delta\vl^4 \\
&=& -2\cos[\varphi] \vl\gamma^{}_1\vl\vl\Delta\vl^2 - 
2\cos[ {\rm arg}(\gamma^{}_{2}) - 2 {\rm arg}(\gamma^{}_{1})  + 2\varphi]\vl\gamma^{}_2\vl\vl\Delta\vl^4,
\end{eqnarray}
where $\varphi = {\rm arg}(\gamma^{}_{1}) + \varphi^{}_1 - \varphi^{}_2$. Following \cite{davydova2022universal} we obtain the current 
\begin{equation}
I[\varphi] = \frac{\delta{\cal F}}{\delta\varphi} = 2\sin[\varphi]\vl\gamma^{}_1\vl\vl\Delta\vl^2 
+ 4\sin[\delta  + 2\varphi]\vl\gamma^{}_2\vl\vl\Delta\vl^4,
\end{equation}
where $\delta = {\rm arg}(\gamma^{}_{2}) - 2 {\rm arg}(\gamma^{}_{1}) $.
The critical current reads 
\begin{equation}
{ \large\vl I\left[\pm\frac{\pi}{2}\right]\large\vl }= 2 \vl\gamma^{}_1\vl\vl\Delta\vl^2 \mp 4\sin[\delta]\vl\gamma^{}_2\vl\vl\Delta\vl^4
\end{equation}
and the current asymmetry
\begin{equation}
\label{eq:Current}
\delta I = - 8\sin[\delta]\vl\gamma^{}_2\vl\vl\Delta\vl^4 .
\end{equation}
Employing Eq.~(\ref{eq:PhasesOfGammas}) we get
\begin{equation}
\delta = {\rm arg}(\gamma^{}_{2}) - 2 {\rm arg}(\gamma^{}_{1}) = \phi[\gamma], 
\end{equation}
where the fourth order phase shift reads
\begin{equation}
\phi[\gamma] = {\rm \arctan}\left[\frac{5}{7}\tan\gamma\right]. 
\end{equation}
Thus
\begin{equation}
\sin[\delta ] = \sin[\phi[\gamma]] = \frac{5}{7}\frac{\tan\gamma}{\sqrt{1 + \frac{25}{49}\tan^2\gamma}} = 
 \frac{5}{g[\gamma]} \sin[\gamma], 
\end{equation}
with $g[\gamma]$ defined below Eqs.~(\textcolor{blue}{13}) in the main text, which is cancelled by the respective term in $\vl\gamma^{}_2\vl$ in  Eq.~(\ref{eq:Current}). 
The current difference depends on the phases only via $\sin[\gamma]=\sin[2\xi^{} - \chi^{}_1 + \chi^{}_2]$. 
Hence, the current difference is 
\begin{equation}
\delta I =-t^2\rho(E_F,B)\frac{|\Delta|^4}{3\Delta_0^4} \sin[2\xi^{} - \chi^{}_1 + \chi^{}_2] .
\end{equation}
If we neglect the phase of the tunneling matrix element $\xi\approx0$, then
\begin{equation}
\label{eq:Asymmetry}
\delta I = -t^2\rho(E_F,B)\frac{|\Delta|^4}{3\Delta_0^4} \sin[\chi^{}_2 - \chi^{}_1] .
\end{equation}
It is convenient to reference the current asymmetry by the density of states, determined in a separate measurement.
In the uniform magnetic field directed along $y$-axis we get 
\begin{equation}
\chi^{}_i = 2\pi\frac{e}{h}\ointop_{{\cal C}^{}_i}d{\bf r}\cdot{\bf A}.
\end{equation}
Using the Stokes theorem and exploiting the translational invariance of the integration measure we then get further
\begin{equation}
\chi^{}_{1/2} = 2\pi B^{}_y\frac{e}{h}\int^H_{0}dz\int^{\pm\lambda^{}_L}_0dx .
\end{equation}
Hence 
\begin{equation}
\delta  = \underbrace{2H\lambda^{}_L}_{=S^{}_{\rm L}}B^{}_y\frac{2\pi}{\Phi^{}_0} = 2\pi \frac{\Phi}{\Phi^{}_0},
\end{equation}
where 
\begin{equation}
\Phi^{}_0 = h/e = 4.136\cdot10^{-15}~{m^2 T} =  4.136\cdot10^{-3} ~{\mu m^2 T}
\end{equation}
denotes the elementary flux quantum, and the flux through the two superconductors
\begin{equation}
\Phi = S^{}_{\rm L}B^{}_y.
\end{equation}
Typical values of the London penetration length in conventional superconductors $\lambda^{}_L$ range from 0.05$\mu m$ to 0.5$\mu m$ (google search).
For the latter value we  estimate the height of the junction to reach for instance the angle $\delta = \pi/4$:
\begin{equation}
\lambda^{}_L H B^{}_y = 0.258\cdot10^{-3}~\mu m^2 T ,
\end{equation}
which gives the height of the superconductor on each side of the junction
\begin{equation}
H = \frac{4.136}{8B^{}_y}\cdot 10^{-3}~\mu mT,
\end{equation}
i.e. for the magnetic field from within $B^{}_y = 0.1T - 2T$ 
\begin{equation}
H \approx 5.2\cdot 10^{-3} \mu m \div 2.6\cdot 10^{-4}\mu m.
\end{equation}
The flux density used in the measurements reported in Ref.~[\onlinecite{pal2022josephson}] was given at $B^{}_y=12\cdot 10^{-3}T$, which gives the height of the junction $H = 4.3\cdot 10^{-2}\mu m$.
With this we can estimate the current asymmetry from Eq.~(\ref{eq:Asymmetry}). Because the Fermi energy is by far the largest energy scale it is justified to approximate $\rho(E_F,B)\sim\rho(E_F)$ for the first insight. 
Since both the DOS and the tunneling amplitude are measurable independently we can reference the current asymmetry  with respect to them. Obviously, the ratio $\vl\Delta\vl/\Delta^{}_0$ can change between 0 and 1. For the first tests we chose $\vl\Delta\vl/\Delta^{}_0\approx 1$, which is true close to the critical temperature. 
Hence, the relative current asymmetry approximately gets
\begin{equation}
\frac{\delta I}{t^2 \rho(E^{}_F)} \approx -\frac{1}{3}\sin[\delta], \;\; \delta = \chi^{}_2-\chi^{}_1 .
\end{equation}
For a uniform magnetic field in $y$-direction $B^{}_y = 0.5 T$ and the height of the superconductor $H = 5\cdot 10^{-3} \mu m$ we get 
\begin{equation}
\delta  = \pi\frac{4H\lambda^{}_LB^{}_y}{\Phi^{}_0} \approx 1.2\pi ,
\end{equation}
and hence, the referenced current asymmetry
\begin{equation}
\frac{\delta I}{t^2 \rho(E^{}_F)} = -\frac{1}{3}\sin(1.2\pi) \approx 0.195.
\end{equation}

\section{Stray magnetic field of the skyrmion}

In case of the thin magnetic films the expression for the stray field can be rewritten in the following form:
%\begin{widetext}
\begin{eqnarray}\label{magnetic field simplified}
B_z(\textbf{r},d)=-\frac{\mu_0M_s}{2}\left(\alpha_z(d)\star\nabla^2 m_z(\textbf{r})+\alpha_\bot(d)\star\vec\nabla\textbf{m}_\bot(\textbf{r})\right), 
\end{eqnarray}
%\end{widetext}
where $\star$ defines convolution in the $x,y$ plane, $\textbf{r}=(x,y)$, and $d$ is the distance between the diode and magnetic texture $\textbf{m}(\textbf{r})$. Coefficients $\alpha_z(d),~\alpha_\bot(d)$ contain information about geometry of the problem. 
The stray field reads
%$\begin{widetext}
\begin{equation}
\label{eq:texture}
\mathbf{B}(\mathbf{\textbf{r}},d)=-\frac{\mu_0M_s}{2}
\left(\begin{array}{ccc}
-\alpha_z(d,l)\frac{\partial ^2}{\partial x^2} &
-\alpha_z(d,l)\frac{\partial ^2}{\partial y \partial x}  & 
 \alpha_{x,y}(d,l)\frac{\partial}{\partial x} \\
-\alpha_z(d,l)\frac{\partial ^2}{\partial y \partial x} & 
-\alpha_z(d,l)\frac{\partial ^2}{\partial y^2}  & 
\alpha_{x,y}(d,l)\frac{\partial}{\partial y}\\
\alpha_{x,y}(d,l)\frac{\partial }{\partial x} & 
\alpha_{x,y}(d,l)\frac{\partial}{\partial y}  & 
\alpha_{z}(d,l)\frac{\partial^2}{\partial x\partial y}
\end{array}
\right)
\left(\begin{array}{c}
m_x(\vec{\mathbf{\textbf{r}}})\\
m_y(\vec{\mathbf{\textbf{r})}}\\
m_z(\vec{\mathbf{\textbf{r}}})
\end{array}
\right)
\end{equation}
%\end{widetext}

where 
\begin{eqnarray}
\alpha_z(d,l)=\frac{l}{2\pi\sqrt{d^2+r^2}},\;\;
\alpha_{x,y}(d,l)=\frac{d\,l}{2\pi(d^2+r^2)^{3/2}}. 
\end{eqnarray}
Here $l$ is the thickness of the film and $r=\sqrt{x^2+y^2}$ is the distance to the source of the stray field (i.e. skyrmion).
Explicitly, the components of the magnetic field are
\begin{eqnarray}
 B^{}_x &=& \frac{\mu_0M_s}{2}\left(\alpha^{}_z \frac{\partial^2}{\partial x^2} m^{}_x + 
 \alpha^{}_z\frac{\partial^2}{\partial x\partial y}m^{}_y - \alpha^{}_{x,y} \frac{\partial}{\partial x}m^{}_z \right) ,\\
 B^{}_y &=& \frac{\mu_0M_s}{2}\left(\alpha^{}_z \frac{\partial}{\partial x\partial y} m^{}_x + \alpha^{}_z\frac{\partial^2}{\partial y^2}m^{}_y - \alpha^{}_{x,y} \frac{\partial}{\partial y}m^{}_z \right) ,\\
 B^{}_z &=& -\frac{\mu_0M_s}{2}\left(
\alpha^{}_{x,y} \frac{\partial}{\partial x} m^{}_x + \alpha^{}_{x,y}\frac{\partial}{\partial y}m^{}_y + \alpha^{}_{z} 
\frac{\partial^2}{\partial x\partial y}m^{}_z 
 \right).
\end{eqnarray}
The phase is obtained as
\begin{equation}
\int_{S} d\vec S\cdot\vec B = \int_{S^{}_x} dS^{}_x B^{}_x + \int_{S^{}_y} dS^{}_y B^{}_y + \int_{S^{}_z} dS^{}_z B^{}_z .
\end{equation}
Since in place of the coordinate $z$ there appears a fixed parameter $d$,  we have $dS^{}_x = 0 = dS^{}_y$.
Consequently, the only contribution to the phase comes from the $B^{}_z$ component 
\begin{equation}
\chi^{}_{1/2} = 2\pi\frac{e}{h}\int^{}_{S^{}_{1/2}}dxdy~B^{}_z(x,y).
\end{equation}
The origin of the coordinate system should be associated with the center of the skyrmion at ($x=0,y=0$). 
For ${x^{}_i,y^{}_i}$ being the position of the nearest corner of each superconducting contact to the center of the skyrmion, $W$ being the width of the contact in $y$-direction, $L$ its length in $x$-direction, the integration limits then are
\begin{equation}
\intop^{x^{}_i+L}_{x^{}_i}dx~\intop^{y^{}_i+W}_{y^{}_i} dy,
\end{equation}
if $W,L\sm \lambda^{}_L$, with the London length needed to be determined from the measuremet.  
If the London length is smaller than the size of the contacts $W,L$, then it must appear in place of them in the integration boundaries.
We notice that ($r=\sqrt{x^2+y^2}$ and $\varphi = \arctan[y/x]$), then for a component of $m$ follows
\begin{eqnarray}
\frac{\partial }{\partial x} m^{}_i(r,\varphi) &=& \frac{\partial r}{\partial x}\frac{\partial m^{}_i}{\partial r} + 
\frac{\partial\varphi}{\partial x}\frac{\partial m^{}_i}{\partial\varphi} = 
\frac{x}{r}\frac{\partial m^{}_i}{\partial r}+ 
\frac{\partial}{\partial x}\arctan\frac{y}{x}\frac{\partial m^{}_i}{\partial\varphi} = 
\frac{x}{r}\frac{\partial m^{}_i}{\partial r} - 
\frac{y}{r^2} \frac{\partial m^{}_i}{\partial\varphi}.
\end{eqnarray}
For each of the components we then explicitly have
\begin{eqnarray}
\label{eq:eqBz1}
\frac{\partial m^{}_x}{\partial x} &=& \frac{x}{r}\frac{\partial m^{}_x}{\partial r} - 
\frac{y}{r^2} \frac{\partial m^{}_x}{\partial\varphi};\\
\label{eq:eqBz2}
\frac{\partial m^{}_y}{\partial y} &=& \frac{y}{r}\frac{\partial m^{}_y}{\partial r} + 
\frac{x}{r^2} \frac{\partial m^{}_y}{\partial\varphi};
\end{eqnarray}
\begin{eqnarray}
\label{eq:eqBz3}
\frac{\partial^2 m^{}_z}{\partial x\partial y} &=& - \frac{xy}{r^3} \frac{\partial m^{}_z}{\partial r}
+ \frac{1}{r^2}\left( 1 - \frac{2xy}{r^2} \right)\frac{\partial m^{}_z}{\partial \varphi} + 
  \frac{y}{r}\left(\frac{x}{r}\frac{\partial^2 m^{}_z}{\partial r^2} - 
  \frac{y}{r^2}\frac{\partial^2 m^{}_z}{\partial r\partial\varphi} \right) + 
  \frac{x}{r^2}\left(\frac{x}{r}\frac{\partial^2 m^{}_z}{\partial r\partial \varphi} - 
  \frac{y}{r^2}\frac{\partial^2 m^{}_z}{\partial\varphi^2} \right).
\end{eqnarray}

%\newpage
\section{Estimation of the DOS due to the coupling to the stray field}

To calculate the density of states we have to consider the geometric series of the following kind:
\begin{equation}
\label{eq:DOS}
\rho(E^{}_F, \vec B) =  \frac{1}{2}{\rm Tr}~\delta(E^{}_F\sigma^{}_0 - h(\vec B)) = \frac{1}{2\pi}{\rm Im}~{\rm Tr}~G(E^{}_F,h(\vec B)),
\end{equation}
where the factor $1/2$ is to account for only one spin projection and the Greens function of the electron gas coupled to the stray tensor reads
\begin{equation}
G^{}_{rr^\prime}(E^{}_F,h(\vec B)) = \langle r\vl\left[E^{}_F\sigma^{}_0  - h(\vec B) - i0^+ \right]^{-1}\vl r^\prime\rangle 
= \langle r\vl\left[G^{-1}_0(E^{}_F)\sigma^{}_0 - \mu^{}_B\vec\sigma\cdot\vec B - i0^+ \right]^{-1}\vl r^\prime\rangle.
\end{equation}

\subsection{The case of the homogeneous external magnetic field}

We start with the homogeneous and coordinate independent magnetic field. 
Then the Greens function is diagonal in the momentum space and becomes
\begin{equation}
\left[G^{-1}_0(E^{}_F)\sigma^{}_0  -  \vec\sigma\cdot\vec B \right]^{-1} = \frac{1}{\det[G^{-1}_0(E^{}_F)\sigma^{}_0  - \mu^{}_B \vec\sigma\cdot\vec B]}
\left[G^{-1}_0(E^{}_F)\sigma^{}_0  +  \mu^{}_B\vec\sigma\cdot\vec B \right],
\end{equation}
where 
\begin{equation}
G^{-1}_0(E^{}_F) = E^{}_F - \epsilon^{}_q - i0^+, \;\; \epsilon^{}_q = \hbar^2q^2/2m,
\end{equation}
and 
\begin{equation}
\det[G^{-1}_0(E^{}_F)\sigma^{}_0  -  \mu^{}_B\vec\sigma\cdot\vec B] = [E^{}_F - \epsilon^{}_q - i0^+ - \mu^{}_B\vl B\vl][E^{}_F - \epsilon^{}_q - i0^+ + \mu^{}_B\vl B\vl]. 
\end{equation}
The density of states of the non-interacting Fermi gas outside of the magnetic field reads
\begin{equation}
\rho^{}_0(E^{}_F) = \frac{1}{\pi}{\rm Im}~{\rm tr}~G^{}_0(E^{}_F) = \int\frac{d^3q}{(2\pi)^3}~\delta(E^{}_F - \epsilon^{}_q) = 
\frac{1}{(2\pi)^2}\left(\frac{2m}{\hbar^2}\right)^{\frac{3}{2}}\sqrt{E^{}_F},
\end{equation}
where in the last expression we restored the SI-units. The density of states thus has the right dimension of inverse volume times inverse energy. After performing the partial fraction decomposition, the density of the states for the homogeneous magnetic field becomes 
\begin{eqnarray}
\rho(E^{}_F, \vec B) &=&  \frac{1}{2\pi}{\rm Im}~\int\frac{d^3q}{(2\pi)^3}~ \left[
\frac{1}{E^{}_F - \epsilon^{}_q - i0^+ - \mu^{}_B\vl B\vl} +
\frac{1}{E^{}_F - \epsilon^{}_q - i0^+ + \mu^{}_B\vl B\vl}
\right] \\
&=&   \frac{1}{2}\int\frac{d^3q}{(2\pi)^3}~ \left[
\delta(E^{}_F - \epsilon^{}_q - \mu^{}_B\vl B\vl) + \delta(E^{}_F - \epsilon^{}_q + \mu^{}_B\vl B\vl)  
\right] \\
&=&   \frac{1}{2} \rho^{}_0(E^{}_F) \left[
\sqrt{1-\mu^{}_B\frac{\vl B\vl}{E^{}_F}} + \sqrt{1+\mu^{}_B\frac{\vl B\vl}{E^{}_F}} 
\right] =  \rho^{}_0(E^{}_F) \left[
1 - \frac{\mu^2_B}{2^22!} \frac{\vl B\vl^2}{E^2_F} - \frac{5!!\mu^4_B}{2^4 4!}\frac{\vl B\vl^4}{E^4_F} - \cdots
\right] \\
\label{eq:DoShomMF}
&=& 
\rho^{}_0(E^{}_F) \left[
1- \sum^{\infty}_{n = 0} \frac{(4n+1)!!}{2^{2n+2} (2n+2)!} \left( \frac{\displaystyle\mu^{}_B\vl B\vl}{E^{}_F} \right)^{2(n+1)}
\right] = \frac{\rho^{}_0(E^{}_F)}{\sqrt{2}}\sqrt{1+\sqrt{1-\mu^2_B\frac{\vl B\vl^2}{E^2_F}}}.
\end{eqnarray}
As function of $\mu^{}_B\vl B\vl/E^{}_F$, the DOS is a smooth curve, equal to $\rho^{}_0(E^{}_F)$
at $\vl B\vl=0$ and $\rho^{}_0(E^{}_F)/\sqrt{2}$ at $\mu^{}_B\vl B\vl=E^{}_F$.

\subsection{The case of the inhomogeneous external magnetic field}

We proceed with the inhomogeneous magnetic field and omit for the time being $\mu^{}_B$ in the subsequent calculation. The expansion in power series in terms of $(\vec\sigma\cdot\vec B^{}_r)^n$ contains only even powers $n$, since the trace over the product of any odd number of Pauli matrices is traceless. We stress that the magnetic field is real. The three-fields contribution is related the topological term. This term turns out to vanish, since the following form is zero
\begin{equation}
 \epsilon^{}_{\alpha\beta\gamma}B^{}_{r\alpha} B^{}_{r\beta} B^{}_{r\gamma} = 0
\end{equation}
by antisymmetry of the tensor $ \epsilon^{}_{\alpha\beta\gamma}$. 
The quadratic term is 
\begin{equation}
\frac{1}{2\pi}{\rm Im}~{\rm tr}~G^{}_0(E^{}_F)[\vec\sigma\cdot\vec B~ G^{}_0(E^{}_F)]^2 = 
\frac{1}{2\pi}{\rm Im}~{\rm Tr}[\sigma^{}_\mu\sigma^{}_\nu]~\sum_{q,p,k}\delta(p+k)B^{}_{p,\mu}B^{}_{k,\nu} ~G^{}_0(E^{}_F,q)G^{}_0(E^{}_F,q+k)G^{}_0(E^{}_F,q),
\end{equation} 
the summation over the doubled indices is understood. The trace over the product of the Pauli matrices is
\begin{equation}
{\rm Tr}[\sigma^{}_\mu\sigma^{}_\nu] = 2\delta^{}_{\mu\nu}.
\end{equation}
To the leading order in gradient expansion we have
\begin{eqnarray}
\frac{1}{2\pi}{\rm Im}~{\rm tr}~G^{}_0(E^{}_F)[\vec\sigma\cdot\vec B~ G^{}_0(E^{}_F)]^2 &\approx& 
\sum_{p,k}\delta(p+k)B^{}_{p,\mu}B^{}_{k,\nu}~\frac{1}{\pi}{\rm Im}~\sum_{q}G^{3}_0(E^{}_F,q)\\
&=& \sum_{r}\vec B^{}_{r}\cdot\vec B^{}_{r} ~ \frac{1}{\pi}{\rm Im}~\int\frac{d^3q}{(2\pi)^3}~\frac{1}{[E^{}_F - \epsilon^{}_q - i0^+]^3} \\
&=& \frac{1}{2}\frac{\partial^2}{\partial E^2_F}~\left(\frac{1}{\pi}{\rm Im}~\int\frac{d^3q}{(2\pi)^3}~\frac{1}{E^{}_F - \epsilon^{}_q - i0^+}\right) \sum_{r}\vec B^{}_{r}\cdot\vec B^{}_{r} \\
&=& \frac{1}{2!}\frac{\partial^2}{\partial E^2_F}~\rho^{}_0(E^{}_F)  \sum_{r}\vec B^{}_{r}\cdot\vec B^{}_{r} 
\end{eqnarray} 
After the evaluation of the derivatives we ultimately have
\begin{equation}
\frac{1}{2\pi}{\rm Im}~{\rm tr}~G^{}_0(E^{}_F)[\vec\sigma\cdot\vec B~ G^{}_0(E^{}_F)]^2 = -\rho^{}_0(E^{}_F) \frac{1}{2^2 2!} 
\frac{1}{E^2_F} \sum_{r}\vec B^{}_{r}\cdot\vec B^{}_{r} .
\end{equation}
The quatric term reads to the leading gradients
\begin{eqnarray}
\nn
\frac{1}{2\pi}{\rm Im}~{\rm tr}~G^{}_0(E^{}_F)[\vec\sigma\cdot\vec B~ G^{}_0(E^{}_F)]^4 &\approx& 
\frac{1}{2\pi}{\rm Im}~{\rm Tr}[\sigma^{}_\mu\sigma^{}_\nu\sigma^{}_\alpha\sigma^{}_\beta]~
\sum_{p,s,t,k}\delta(p+k+s+t)B^{}_{p,\mu}B^{}_{k,\nu}B^{}_{s,\alpha}B^{}_{t,\beta}~\sum_{q}G^{5}_0(E^{}_F,q)\\
&=& 
\frac{1}{2\pi}{\rm Im}~{\rm Tr}[\sigma^{}_\mu\sigma^{}_\nu\sigma^{}_\alpha\sigma^{}_\beta]~
\sum_{r}B^{}_{r,\mu}B^{}_{r,\nu}B^{}_{r,\alpha}B^{}_{r,\beta}~\sum_{q}G^{5}_0(E^{}_F,q)
\end{eqnarray} 
i.e. the four-field correlation term is local. The trace is evaluated as 
\begin{equation}
\frac{1}{2}~{\rm Tr}[\sigma^{}_\mu\sigma^{}_\nu\sigma^{}_\alpha\sigma^{}_\beta] =
\delta^{}_{\alpha\beta}\delta^{}_{\mu\nu} +  \delta^{}_{\alpha\nu}\delta^{}_{\beta\mu} - \delta^{}_{\alpha\mu}\delta^{}_{\beta\nu},
\end{equation}
i.e. only one contribution survives upon summing out the Pauli indices:
\begin{equation}
\frac{1}{2}~{\rm Tr}[\sigma^{}_\mu\sigma^{}_\nu\sigma^{}_\alpha\sigma^{}_\beta]~
\sum_{r}B^{}_{r,\mu}B^{}_{r,\nu}B^{}_{r,\alpha}B^{}_{r,\beta} = \sum_{r} \left(\vec B^{}_{r}\cdot\vec B^{}_{r}  \right)^2 .
\end{equation}
The amplitude is evaluated similarly to the quadratic term:
\begin{equation}
\frac{1}{\pi}{\rm Im}~\sum_{q}G^{5}_0(E^{}_F,q) = \frac{1}{4!}\frac{\partial^4}{\partial E^4_F}~\rho^{}_0(E^{}_F) = 
- \rho^{}_0(E^{}_F) \frac{5!!}{2^4 4!} \frac{1}{E^4_F},
\end{equation}
where we use the short hand $5!!=1\cdot3\cdot5$, i.e. as a whole for this contribution 
\begin{equation}
\frac{1}{2\pi}{\rm Im}~{\rm tr}~G^{}_0(E^{}_F)[\vec\sigma\cdot\vec B~ G^{}_0(E^{}_F)]^4 \approx  
- \rho^{}_0(E^{}_F) \frac{5!!}{2^4 4!} \frac{1}{E^4_F} ~ \sum_{r} \left(\vec B^{}_{r}\cdot\vec B^{}_{r}  \right)^2.
\end{equation}
Restoring again the Bohr magneton in the formulae we deduce the generic form of the element of the infinite series 
\begin{equation}
\frac{1}{2\pi}{\rm Im}~{\rm tr}~G^{}_0(E^{}_F)[\vec\sigma\cdot\vec B~ G^{}_0(E^{}_F)]^{2n} \approx 
- \rho^{}_0(E^{}_F)  \frac{(4n+1)!!}{2^{2n+2} (2n+2)!} 
\sum_r \left(\mu^2_B\frac{\displaystyle\vec B^{}_{r}\cdot\vec B^{}_{r}}{E^2_F} \right)^{n+1}, \;\; n = 0,1,2,3\dots,
\end{equation}
where we recognize the same combinatorics of the series element as in the case with the homogeneous magnetic field Eq.~(\ref{eq:DoShomMF}). 
Hence, the transition from this case to the case of uniform fields goes straightforwardly by omitting the averaging over the position space. Formally, the sum over $n$ converges, giving 
\begin{equation}
\label{eq:DOS-B}
\rho(E^{}_F, \vec B) = \rho^{}_0(E^{}_F) \left( 1 -  
\sum_{\textbf{r}} \left[1 - \frac{1}{\sqrt{2}} \sqrt{\displaystyle 1 + \sqrt{1 - \frac{\mu^2_B}{E^2_F}\displaystyle\vec B(\textbf{r})\cdot\vec B(\textbf{r}) }} \right]
\right).
\end{equation}
%Summation over $\textbf{r}$ in Eq.(\ref{eq:DOS-B}) in the coarse-graining approach denotes the summation over 
%the magnetic texture forming the field. 

%\bibliography{2nems.bib}
\bibliographystyle{apsrev4-1}